\pdfoutput=1
\documentclass[11pt]{article}
\usepackage[]{acl}
\usepackage{times}
\usepackage{latexsym}
\usepackage[T1]{fontenc}
\usepackage[utf8]{inputenc}
\usepackage{microtype}
\usepackage{booktabs} 
\usepackage{tabularx} 
\usepackage{hyperref}       
\usepackage{url}            
\usepackage{xurl} 
\usepackage{amsfonts}       
\usepackage{nicefrac}       
\usepackage{microtype}      
\usepackage{xcolor}         
\usepackage{graphicx}       
\usepackage{subfig} 
\usepackage{amsmath}
\usepackage{algorithm}
\usepackage{listings}
\usepackage{pdflscape} 
\usepackage{changepage} 

\title{Jasper and Stella: distillation of SOTA embedding models}

\author{
 \textbf{Dun Zhang\textsuperscript{1}},
 \textbf{Jiacheng Li\textsuperscript{1}\thanks{$^{\ast}$Dun Zhang and Jiacheng Li make equal contributions to this
work.}},
 \textbf{Ziyang Zeng\textsuperscript{1,2}},
 \textbf{Fulong Wang\textsuperscript{1}}
\\
 \textsuperscript{1}NovaSearch Team \\
 \textsuperscript{2}Beijing University of Posts and Telecommunications
\\
\small{
 \texttt{infgrad@163.com} \quad
 \texttt{jcli.nlp@gmail.com}
}
\\
\small{
 \texttt{ziyang1060@bupt.edu.cn} \quad
 \texttt{wangfl1989@163.com}
}
}

\begin{document}
\maketitle
\begin{abstract}


A crucial component in many deep learning applications, such as Frequently Asked Questions (FAQ) and Retrieval-Augmented Generation (RAG), is dense retrieval. 
In this process, embedding models transform raw text into numerical vectors. 
However, the embedding models that currently excel on text embedding benchmarks, like the Massive Text Embedding Benchmark (MTEB), often have numerous parameters and high vector dimensionality. 
This poses challenges for their application in real-world scenarios. 
To address this issue, we propose a novel multi-stage distillation framework that enables a smaller student embedding model to distill multiple larger teacher embedding models through three carefully designed losses.
Meanwhile, we utilize Matryoshka Representation Learning (MRL) to reduce the vector dimensionality of the student embedding model effectively.
Our student model named Jasper with 2 billion parameters, built upon the Stella embedding model, obtained the No.3 position on the MTEB leaderboard (as of December 24, 2024), achieving average 71.54 score across 56 datasets.
We have released the model and data on the Hugging Face Hub \footnote{\url{https://huggingface.co/infgrad/jasper_en_vision_language_v1}} \footnote{\url{https://huggingface.co/datasets/infgrad/jasper_text_distill_dataset}}, and the training codes are available in this project repository \footnote{\url{https://github.com/NLPJCL/RAG-Retrieval}}.
\end{abstract}

\section{Introduction}
With the rapid development of natural language processing technologies, text embedding models play a crucial role in text representation \cite{embedding_representations}, information retrieval \cite{DenseTextRetrieval}, and text generation tasks \cite{rag}.
By mapping words, sentences, or documents into a high-dimensional continuous space, these models bring similar texts closer together in their vector representations, thereby not only enhancing the manipulability of textual data but also significantly improving the performance of various downstream tasks \cite{search_www,search_rag,search_multimodal}.

However, embedding models that demonstrate excellent performance on the METB leaderboard\footnote{\url{https://huggingface.co/spaces/mteb/leaderboard}} \cite{MTEB} usually contain a large number of parameters and high vector dimensions.
For instance, both NV-Embed-v2 \cite{lee2024nv,moreira2024nv} and bge-en-icl \cite{bge_embedding,li2024makingtextembeddersfewshot} have 7 billion parameters and 4096-dimensional vector representations.
These characteristics lead to slow inference speeds and high storage costs, posing a significant challenge to their direct practical application.

To address the aforementioned challenges, we propose a novel multi-stage knowledge distillation framework for embedding models. Knowledge distillation is widely recognized for enhancing the effectiveness of dense retrieval training \cite{distill_colbert, distll2}. 
In our framework, we introduce three carefully designed loss functions to distill knowledge from the teacher model to the student model. 
These loss functions shift from a specific constraint to a broader constraint.
The first, cosine loss, calculates the absolute difference in text representations between the student and teacher models. 
The pointwise signal derived from a single text is straightforward, yet its limited optimization direction tends to readily lead to overfitting on the training data.
Thus, we introduce the similarity loss, which measures the semantic discrepancies between the student and teacher models from a text-pair perspective.
Additionally, we design the relative similarity distillation loss to further leverage relative ranking information. 
This ensures that the student model learns the teacher’s ranking preferences across all potential positive and negative text pairs within the batch, thereby improving the robustness of embedding learning.

To further improve the performance of the student model, we utilize multiple powerful large embedding models as teachers. Specifically, we concatenate the vectors produced by all teacher models to create the final ground truth, which inevitably leads to an increase in the student model's vector dimension.
To address this issue, we adopt a Matryoshka Representation Learning (MRL)-based training method \cite{kusupati2024matryoshkarepresentationlearning} to effectively compress the student model's vector representation.
Additionally, to develop the multimodal retrieval capability of our student model, we integrate a vision encoder and introduce a self-distillation mechanism to align the visual embeddings with the textual embeddings.
In terms of the overall training process, we employ a 4-stage distillation approach to progressively transfer knowledge from the teacher models to the student model. 
Each stage focuses on specific aspects, combining three loss functions and fine-tuning different parameters of the student model to ensure a smooth and effective distillation process.

Experimental results on the MTEB leaderboard demonstrate that our student model named Jasper with 2 billion (2B) parameters, primarily built upon the foundation of the Stella embedding model, delivers excellent performance (average 71.54 score across 56 datasets) comparable to other embedding models with 7 billion (7B) parameters, and significantly outperforms models with fewer than 2B parameters.

The main contributions of this paper can be summarized as follows:
\begin{itemize}
    \item [(1)] We propose a novel multi-stage distillation framework, which enables a smaller student embedding model to effectively distill knowledge from multiple larger teacher embedding models through three carefully designed loss functions.
    \item [(2)] Our 2B Jasper model obtained the No.3 position on the MTEB leaderboard (as of December 24, 2024), producing results comparable to other top-ranked 7B embedding models and significantly outperforming other models with less than 2B parameters.
\end{itemize}

\section{Methods}

\subsection{Definitions}
For a more comprehensive introduction of our model and distillation framework, we make the following definitions:

\begin{itemize}
    \item Student Model: The text embedding model that is the subject of training, tasked with learning to produce effective vector representations.
    \item Teacher Model: The state-of-the-art (SOTA) embedding model serving as a teacher, guiding the student model in generating effective vectors. Notably, the teacher model will not be trained.
    \item $s_x$:  The normalized vector representation of a text $x$ produced by the student model.
    \item  $t_x$: The vector representation of the same text $x$, first normalized, then concatenated, and normalized again, produced by multiple teacher models.
    \item  $S_X$: A matrix of normalized vector representations for a batch of text $X$ produced by the student model.
    \item  $T_X$: A corresponding matrix of vector representations for the same batch of text $X$, first normalized, then concatenated, and subsequently normalized again, generated by multiple teacher models.
\end{itemize}

\subsection{Model Architecture} 

\begin{figure}[th]
    \centering
    \begin{center}
        \includegraphics[width=0.45\textwidth]{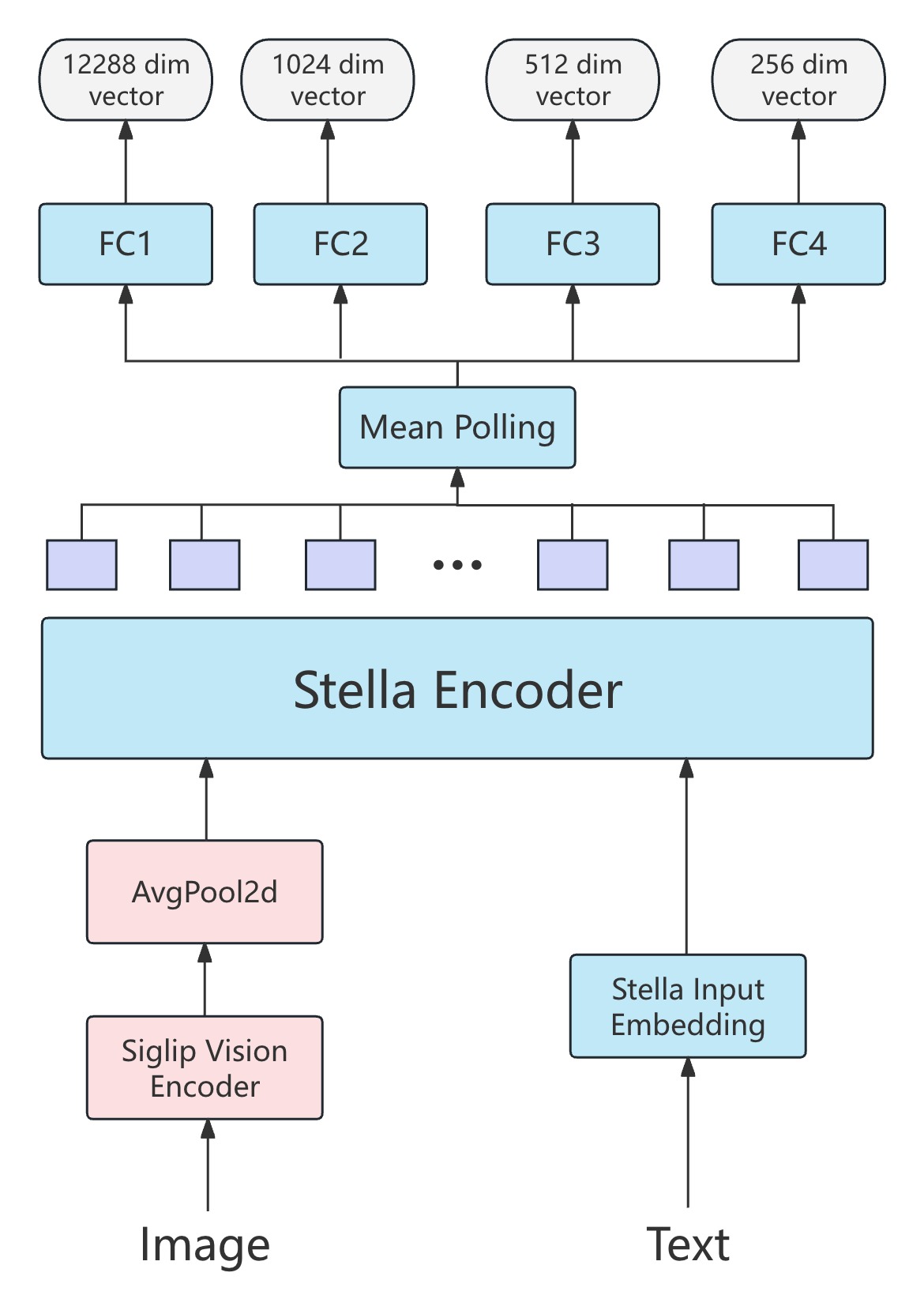}
        \caption{The model architecture of Jasper model.}
        \label{fig:model_architecture}
    \end{center}
\end{figure}

Our student model architecture follows the simple and standard design of combining a language model with a vision encoder. 
As shown in Figure~\ref{fig:model_architecture}, it consists of four components:
\begin{enumerate}
    \item A encoder-based language model that generates text embeddings through mean pooling.
    \item A vision encoder that independently maps images into vision token embeddings.
    \item A pooler that maps vision token embeddings to the same dimension as the language model’s input textual embeddings, while reducing the length of visual token sequences.
    \item Several fully connected (FC) layers that project the embeddings to a specific dimension for the final output.
\end{enumerate}

\subsection{ Stage 1\&2: Distillation from Multiple Teachers}

In the first two stages of distillation, we use a fully connected layer to map the vectors of the student model onto the dimensions of the teacher models. 
Specifically, we employ \textit{NV-Embed-v2}\footnote{\url{https://huggingface.co/nvidia/NV-Embed-v2}} and \textit{stella\_en\_1.5B\_v5}\footnote{\url{https://huggingface.co/dunzhang/stella_en_1.5B_v5}} as teacher models, which have vector dimensions of 4096 and 8192, respectively. 
After the mapping process, the student model's vector dimension is adjusted to 12288, equal to the combined vector dimensions of two teacher models (4096 + 8192).

The objective of the first two stages is to enable the student model to effectively learn text representations from multiple teacher models by aligning its output vectors with the corresponding teacher vectors.
To achieve this goal, we carefully design three loss functions that progress from a specific to a broader perspective.
The first loss function is cosine loss, which is formulated as follows:
\begin{equation}
\label{loss1}
\mathcal{L}_{cosine}=  \sum_x 1 - s_x \cdot t_x.
\end{equation}
The $\mathcal{L}_{\text{cosine}}$ is designed to minimize the angular difference between student and teacher vectors in the high-dimensional space, with the aim of aligning their absolute text representations. 
However, the $\mathcal{L}_{cosine}$ value generally does not converge to zero, suggesting a persistent angular discrepancy between the student and the teachers.
Meanwhile, the pointwise signal derived from a single text has a limited optimization direction, which can easily lead to overfitting on the training data.

\begin{equation}
\label{loss2}
\mathcal{L}_{sim} = MSE(S_{X}S_{X}^T, T_{X}T_{X}^T))
\end{equation}

To complement the limitations of $\mathcal{L}_{cosine}$, we introduce the second loss function, similarity loss, as defined in \eqref{loss2}, which models the semantic matching differences between the student and teacher models from a text-pair perspective. 
This loss function ensures a relatively consistent judgment of similarity between the student model and the teacher models, without enforcing an absolute fit between the student model and the teacher model.

\begin{equation}
\label{loss3}
\begin{split}
 \mathcal{L}_{resim} = \frac{1}{N} \sum_{t_i \cdot t_j > t_m \cdot t_n} MAX(0,\\ s_m \cdot s_n - s_i \cdot s_j + margin)
 \end{split}
\end{equation}


To further leverage relative comparison signals, inspired by CoSENT loss\footnote{\url{https://spaces.ac.cn/archives/8847}}, we propose the third loss function, relative similarity distillation loss, as defined in \eqref{loss3}. 
For each batch of text data, we employ teacher models to automatically generate soft labels for all text pairs, thereby identifying potential positive and negative samples. Subsequently, the student model is trained to ensure that the similarity between positive pairs exceeds that between negative pairs, with the $margin$ hyperparameter controlling the degree of this difference.
If the batch size is $m$, the total number of text pairs (\textit{i.e.}, $N$) is given by $C_{C_{m}^{2}}^{2}$.

\begin{equation}
 \mathcal{L} = \lambda_{1}\mathcal{L}_{cosine}+\lambda_{2}\mathcal{L}_{sim}+\lambda_{3}\mathcal{L}_{resim}
\end{equation}

The final loss $\mathcal{L}$ is a weighted sum of the aforementioned three loss functions. where $\lambda_{1}$,$\lambda_{2}$, and $\lambda_{3}$ are hyperparameters. 
The biggest advantage of distillation vectors is that we do not need any supervised data. 
Without considering resource constraints, we can use trillions of unsupervised texts for distillation training to achieve extreme performance for a given model size.

Notably, the main difference between stage 1 and stage 2 lies in the trained parameters. 
In stage 1, only the fully connected layer (FC1) is trained, whereas in stage 2, both the fully connected layer (FC1) and the last three encoder layers of the student model are trained.

\subsection{Stage 3: Dimension Reduction}

In the first two stages, the student model is trained by learning from the teacher models. 
Specifically, we concatenate the vectors produced by the two teacher models, resulting in a student model vector with a dimensionality of 12,288 (4,096 + 8,192), which is impractically large. 
Inspired by MRL \cite{kusupati2024matryoshkarepresentationlearning}, we introduce three additional, independent fully connected layers (FC2, FC3, and FC4) to generate low-dimensionality vectors, each achieving a different level of dimension reduction. 
For instance, by incorporating the fully connected layer FC3 with a shape of (1536\footnote{This refers to the dimensionality of the encoder layer's hidden state.}, 512), we obtain a more manageable 512-dimensional vector space.

For the three FC layers, since the dimensions of the reduced vectors do not align with those of the concatenated teacher vector, the $\mathcal{L}_{cosine}$ is omitted and only the $\mathcal{L}_{sim}$ and $\mathcal{L}_{resim}$ are utilized.
To ensure the accuracy of the vectors generated from the FC1 layer (\textit{i.e.}, the 12288-dimensional vectors), they continue to be trained using all three loss functions.
During this stage, all parameters of the student model are trained.

In addition to the previously mentioned dimension reduction method, we present a potentially promising approach to self-distillation, where the aligned vectors from an earlier stage of the student model's training serve as teacher vectors.
Specifically, we propose to utilize the 12288-dimensional vectors output from the FC1 layer to serve as teachers for the shorter vectors generated by the other three FC layers. 
This approach provides a unique advantage by enabling the reduction of the dimensionality of any embedding model, utilizing only unsupervised data and the model itself.
Given that this paper primarily focuses on introducing the training methods of the Stella and Jasper models, we did not conduct experiments to evaluate the specific merits of this proposed approach.

\subsection{ Stage 4: Unlock Multimodal Potential }
In stage 4, we leverage image-caption pairs as the training dataset, focusing exclusively on training the visual encoder while keeping the other components frozen. 
The training process is based on self-distillation, where the caption’s vector representation serves as the teacher vector, and the image’s vector representation acts as the student vector. 
All fully connected layers introduced in previous stages are employed to generate multiple pairs of student and teacher vectors. 
For each pair, we calculate three losses, which are then averaged to obtain the final loss.

It is important to note that this stage achieves only a preliminary alignment between the text and image modalities, leaving significant room for improvement. 
In future work, we aim to further explore and refine the modality alignment process.

\section{Experiments}

\begin{table*}[!t]
    \centering
    \scriptsize
    \resizebox{\textwidth}{!}{\begin{tabular}{lccccccccc}
    \toprule
        Model &Model Size & Average(56 datasets) & Classification & Clustering & PairClassification & Reranking & Retrieval & STS & Summarization \\ 
        \midrule
        NV-Embed-v2 &7851M & 72.31 & 90.37 & 58.46 & 88.67 & 60.65 & 62.65 & 84.31 & 30.7 \\ 
        bge-en-icl  &7111M & 71.67 & 88.95 & 57.89 & 88.14 & 59.86 & 62.16 & 84.24 & 30.77 \\ 
        Stella\_en\_1.5B\_v5 &1543M & 71.19 & 87.63 & 57.69 & 88.07 & 61.21 & 61.01 & 84.51 & 31.49 \\
        SFR-Embedding-2\_R &7111M & 70.31 &89.05 &56.17 &88.07 &60.14 &60.18 &81.26 &30.71 \\
        \hline
        gte-Qwen2-1.5B-instruct &1776M & 67.16 & 82.47 & 48.75 & 87.51 & 59.98 & 58.29 & 82.73 & 31.17 \\ 
        voyage-lite-02-instruct &1220M &67.13 &79.25 &52.42 &86.87 &58.24 &56.60 &85.79 &31.01 \\
        Jasper (our model) & 1543M+400M & 71.54 & 88.49 & 58.04 & 88.07 &  60.91 &  61.33 & 84.67 & 31.42 \\ 
        \bottomrule
    \end{tabular}}
    \caption{ MTEB Results as of December 24, 2024. We use the original model names on the leaderboard for clarity.}
    \label{tab:mteb_results}
\end{table*}

\subsection{Implementation details}
Our model named Jasper is initialized from \textit{stella\_en\_1.5B\_v5} and \textit{google/siglip-so400m-patch14-384} \cite{zhai2023sigmoidlosslanguageimage, alabdulmohsin2024gettingvitshapescaling}. 
\textit{stella\_en\_1.5B\_v5} and \textit{NV-Embed-v2} are our teacher models. 
The total number of parameters in our Jasper model is 1.9B (stella 1.5B parameters and siglip 400M parameters). 
For hyperparameters, we set
$\lambda_1$ = 10, $\lambda_2$ = 200, $\lambda_3$ = 20, margin = 0.015.

In all four stages, the model is trained using 8 × RTX A6000 GPUs, with a maximum input length of 512 tokens, mixed precision training (BF16), DeepSpeed ZERO-stage-2, and the AdamW optimizer.
During stage 1 (distillation training), the batch size is set to 128, the learning rate is 1e-4 per step, and the model checkpoint at step 4000 is selected as the final model.
In the case of stage 2 (also distillation training), the batch size remains 128, the learning rate drops to 8e-5 per step, and the final model is the checkpoint at step 7000.
For stage 3 (dimension reduction training), the batch size is again 128, the learning rate is adjusted to 7e-5 per step, and the checkpoint at step 2200 serves as the final model.
Lastly, in stage 4 (multimodal training), the batch size is reduced to 90, the learning rate returns to 1e-4 per step, and the final model is chosen from the checkpoint at step 3500.

\subsection{Datasets}

In stage 1, stage 2 and stage 3, we use  \textit{fineweb-edu} \cite{lozhkov2024fineweb-edu} as our main text training dataset, which accounts for 80\% of the full text data.
The remaining 20\% of the text data comes from \textit{sentence-transformers/embedding-training-data}\footnote{\url{https://huggingface.co/datasets/sentence-transformers/embedding-training-data}}. 
The reason we choose the \textit{sentence-transformers/embedding-training-data} is that the majority of the \textit{fineweb-edu} data consists of passages. 
However, in addition to passages, we also require questions to enhance the diversity of our training data.
The total amount of text training data is 8 million.

For the documents in our dataset, we perform the following actions:
\begin{enumerate}
\item We randomly select 30\% of the documents and divide them into short texts, each consisting of 1 to 10 sentences.
\item We randomly select 0.08\% of the text and shuffle the words within it.
\end{enumerate}

In stage 4, we use the caption data of \textit{BAAI/Infinity-MM} \cite{gu2024infinitymmscalingmultimodalperformance} as our vision training data.

\subsection{Results}
We evaluate the proposed Jasper and Stella models on the full MTEB benchmark, which encompasses 15 retrieval datasets, 4 reranking datasets, 12 classification datasets, 11 clustering datasets, 3 pair classification datasets, 10 semantic textual similarity datasets, and 1 summarization dataset.

Table \ref{tab:mteb_results} presents the average score of our Jasper model across the overall performance and seven subcategory tasks of the METB benchmark. We compare our model with other frontier models on the MTEB leaderboard, as well as those with fewer than 2B parameters. 
Experimental results demonstrate that our Jasper model significantly outperforms other models with fewer than 2B parameters. 
Furthermore, despite having only 2B parameters, our model produces results that are comparable to those of models with 7B parameters.

\section{Discussion}

\subsection{Instruction Robustness}
Instruction-based embedding models require an instruction to be prepended to a query or passage during text encoding. 
Currently, many state-of-the-art text embedding models use instructions to prompt the model and obtain better embeddings. 
Similar to the usage of large language models  \cite{llm}, different tasks necessitate different instructions, which is both logical and intuitive. 
Therefore, the ability to understand instructions is crucial for these text embedding models.

Jasper is also an instruction-based embedding model. 
To demonstrate the impact of different prompts on the Jasper model, we conducted a simple experiment. 
Specifically, we evaluated Jasper on some short evaluation tasks using similar instructions generated by GPT-4o. 
Table \ref{tab:mteb_inst} lists all the original and modified instructions.
Based on the results shown in Table \ref{tab:mteb_inst_results}, we conclude that our Jasper model is robust to instructions and can accurately understand different instructions.

\begin{table*}[!t]
    \centering
    \scriptsize
    \resizebox{\textwidth}{!}{\begin{tabular}{ll}
    \toprule
        Original Instruction & Synonym of Original Instruction \\   
        \midrule
        Classify the sentiment expressed in the given movie review text from the IMDB dataset & Determine the sentiment conveyed in the provided movie review text from the IMDB dataset. \\
        Identify the topic or theme of StackExchange posts based on the titles & Determine the subject or theme of StackExchange posts based on the titles. \\
        Given a news summary, retrieve other semantically similar summaries & Given a news summary, find other summaries with similar meanings. \\
        Retrieve duplicate questions from StackOverflow forum & Find duplicate questions on the StackOverflow forum. \\
        Given a title of a scientific paper, retrieve the titles of other relevant papers & Given the title of a scientific paper, find the titles of other related papers. \\
        Classify the sentiment of a given tweet as either positive, negative, or neutral & Determine the sentiment of a given tweet as positive, negative, or neutral. \\
        Given a claim, find documents that refute the claim & Given a claim, locate documents that contradict the claim. \\
        Given a question, retrieve relevant documents that best answer the question & Given a question, find relevant documents that best answer it. \\
        Retrieve tweets that are semantically similar to the given tweet & Find tweets that have similar meanings to the given tweet. \\
        Retrieve semantically similar text. & Find text with similar meanings. \\
        Identify the main category of Medrxiv papers based on the titles & Determine the primary category of Medrxiv papers based on the titles. \\
        Retrieve duplicate questions from AskUbuntu forum & Find duplicate questions on the AskUbuntu forum. \\
        Given a question, retrieve detailed question descriptions from Stackexchange that are duplicates to the given question & Given a question, find detailed question descriptions from Stackexchange that are duplicates. \\
        Identify the main category of Biorxiv papers based on the titles and abstracts & Determine the primary category of Biorxiv papers based on the titles and abstracts. \\
        Given a financial question, retrieve user replies that best answer the question & Given a financial question, find user replies that best answer it. \\
        Given a online banking query, find the corresponding intents & Given an online banking query, identify the corresponding intents. \\
        Identify the topic or theme of the given news articles & Determine the subject or theme of the given news articles. \\
        Classify the emotion expressed in the given Twitter message into one of the six emotions: anger, fear, joy, love, sadness, and surprise & Determine the emotion expressed in the given Twitter message as one of six emotions: anger, fear, joy, love, sadness, and surprise. \\
        Given a user utterance as query, find the user intents & Given a user utterance as a query, identify the user intents. \\
        Identify the main category of Biorxiv papers based on the titles & Determine the primary category of Biorxiv papers based on the titles. \\
        Classify the given Amazon review into its appropriate rating category & Classify the given Amazon review into its appropriate rating category. \\
        Given a scientific claim, retrieve documents that support or refute the claim & Given a scientific claim, find documents that support or contradict the claim. \\
        Identify the topic or theme of StackExchange posts based on the given paragraphs & Determine the subject or theme of StackExchange posts based on the given paragraphs. \\
        Given a scientific paper title, retrieve paper abstracts that are cited by the given paper & Given a scientific paper title, find paper abstracts that are cited by the given paper. \\
        Classify the given comments as either toxic or not toxic & Classify the given comments as toxic or non-toxic. \\
        Classify the intent domain of the given utterance in task-oriented conversation & Determine the intent domain of the given utterance in task-oriented conversation. \\
        Retrieve duplicate questions from Sprint forum & Find duplicate questions on the Sprint forum. \\
        Given a user utterance as query, find the user scenarios & Given a user utterance as a query, identify the user scenarios. \\
        Classify the intent of the given utterance in task-oriented conversation & Determine the intent of the given utterance in task-oriented conversation. \\
        Classify a given Amazon customer review text as either counterfactual or not-counterfactual & Classify a given Amazon customer review text as counterfactual or non-counterfactual. \\
        Identify the main category of Medrxiv papers based on the titles and abstracts & Determine the primary category of Medrxiv papers based on the titles and abstracts. \\
        Given a query on COVID-19, retrieve documents that answer the query & Given a query on COVID-19, find documents that answer the query. \\
        \bottomrule
    \end{tabular}}
    \caption{Original instructions and corresponding synonyms.}
    \label{tab:mteb_inst}
\end{table*}

\begin{table}[!t]
    \centering
    \scriptsize
    \resizebox{0.45\textwidth}{!}{\begin{tabular}{llcc}
    \toprule
        \textbf{Task Type} & \textbf{Task Name} & \textbf{Original Score} & \textbf{Score with Modified Instructions} \\ 
        \midrule
\textbf{Classification}     & \textbf{MTOPDomainClassification}               & 0.992                    & 0.992                                        \\
\textbf{Classification}     & \textbf{AmazonCounterfactualClassification}     & 0.958                    & 0.957                                        \\
\textbf{Classification}     & \textbf{TweetSentimentExtractionClassification} & 0.773                    & 0.776                                        \\
\textbf{Classification}     & \textbf{EmotionClassification}                  & 0.877                    & 0.859                                        \\
\textbf{Classification}     & \textbf{MassiveIntentClassification}            & 0.853                    & 0.854                                        \\
\textbf{Classification}     & \textbf{AmazonReviewsClassification}            & 0.629                    & 0.630                                        \\
\textbf{Classification}     & \textbf{MassiveScenarioClassification}          & 0.912                    & 0.912                                        \\
\textbf{Classification}     & \textbf{Banking77Classification}                & 0.873                    & 0.875                                        \\
\textbf{Classification}     & \textbf{ImdbClassification}                     & 0.971                    & 0.971                                        \\
\textbf{Classification}     & \textbf{ToxicConversationsClassification}       & 0.913                    & 0.910                                        \\
\textbf{Classification}     & \textbf{MTOPIntentClassification}               & 0.915                    & 0.912                                        \\
\textbf{Clustering}         & \textbf{MedrxivClusteringS2S}                   & 0.448                    & 0.448                                        \\
\textbf{Clustering}         & \textbf{StackExchangeClusteringP2P}             & 0.494                    & 0.492                                        \\
\textbf{Clustering}         & \textbf{StackExchangeClustering}                & 0.800                    & 0.795                                        \\
\textbf{Clustering}         & \textbf{TwentyNewsgroupsClustering}             & 0.630                    & 0.625                                        \\
\textbf{Clustering}         & \textbf{MedrxivClusteringP2P}                   & 0.470                    & 0.468                                        \\
\textbf{Clustering}         & \textbf{BiorxivClusteringS2S}                   & 0.476                    & 0.475                                        \\
\textbf{Clustering}         & \textbf{BiorxivClusteringP2P}                   & 0.520                    & 0.518                                        \\
\textbf{PairClassification} & \textbf{TwitterURLCorpus}                       & 0.877                    & 0.877                                        \\
\textbf{PairClassification} & \textbf{SprintDuplicateQuestions}               & 0.964                    & 0.964                                        \\
\textbf{PairClassification} & \textbf{TwitterSemEval2015}                     & 0.803                    & 0.801                                        \\
\textbf{Reranking}          & \textbf{StackOverflowDupQuestions}              & 0.546                    & 0.548                                        \\
\textbf{Reranking}          & \textbf{SciDocsRR}                              & 0.891                    & 0.890                                        \\
\textbf{Reranking}          & \textbf{AskUbuntuDupQuestions}                  & 0.674                    & 0.676                                        \\
\textbf{Retrieval}          & \textbf{CQADupstackMathematicaRetrieval}        & 0.369                    & 0.370                                        \\
\textbf{Retrieval}          & \textbf{CQADupstackStatsRetrieval}              & 0.413                    & 0.413                                        \\
\textbf{Retrieval}          & \textbf{CQADupstackTexRetrieval}                & 0.362                    & 0.362                                        \\
\textbf{Retrieval}          & \textbf{SCIDOCS}                                & 0.247                    & 0.247                                        \\
\textbf{Retrieval}          & \textbf{CQADupstackEnglishRetrieval}            & 0.543                    & 0.543                                        \\
\textbf{Retrieval}          & \textbf{ArguAna}                                & 0.653                    & 0.652                                        \\
\textbf{Retrieval}          & \textbf{TRECCOVID}                              & 0.865                    & 0.866                                        \\
\textbf{Retrieval}          & \textbf{CQADupstackUnixRetrieval}               & 0.482                    & 0.482                                        \\
\textbf{Retrieval}          & \textbf{CQADupstackGamingRetrieval}             & 0.632                    & 0.633                                        \\
\textbf{Retrieval}          & \textbf{CQADupstackGisRetrieval}                & 0.444                    & 0.448                                        \\
\textbf{Retrieval}          & \textbf{CQADupstackWordpressRetrieval}          & 0.388                    & 0.386                                        \\
\textbf{Retrieval}          & \textbf{FiQA2018}                               & 0.601                    & 0.601                                        \\
\textbf{Retrieval}          & \textbf{SciFact}                                & 0.805                    & 0.805                                        \\
\textbf{Retrieval}          & \textbf{CQADupstackPhysicsRetrieval}            & 0.549                    & 0.548                                        \\
\textbf{Retrieval}          & \textbf{NFCorpus}                               & 0.431                    & 0.431                                        \\
\textbf{Retrieval}          & \textbf{CQADupstackProgrammersRetrieval}        & 0.505                    & 0.505                                        \\
\textbf{Retrieval}          & \textbf{CQADupstackAndroidRetrieval}            & 0.571                    & 0.571                                        \\
\textbf{Retrieval}          & \textbf{CQADupstackWebmastersRetrieval}         & 0.464                    & 0.464                                        \\
\textbf{STS}                & \textbf{BIOSSES}                                & 0.848                    & 0.854                                        \\
\textbf{STS}                & \textbf{STS13}                                  & 0.897                    & 0.888                                        \\
\textbf{STS}                & \textbf{STS12}                                  & 0.803                    & 0.804                                        \\
\textbf{STS}                & \textbf{STSBenchmark}                           & 0.888                    & 0.886                                        \\
\textbf{STS}                & \textbf{STS15}                                  & 0.902                    & 0.900                                        \\
\textbf{STS}                & \textbf{STS14}                                  & 0.853                    & 0.851                                        \\
\textbf{STS}                & \textbf{STS16}                                  & 0.864                    & 0.869                                        \\
\textbf{STS}                & \textbf{STS22}                                  & 0.672                    & 0.748                                        \\
\textbf{STS}                & \textbf{SICK-R}                                 & 0.822                    & 0.823                                        \\
\textbf{STS}                & \textbf{STS17}                                  & 0.911                    & 0.908                                        \\
\textbf{Summarization}      & \textbf{SummEval}                               & 0.313                    & 0.314                                        \\
\midrule
\multicolumn{2}{l}{\textbf{Average Score}}                                             & 0.686                    & 0.687                                       \\

    \bottomrule
    \end{tabular}}
    \caption{MTEB Results on different instructions.}
    \label{tab:mteb_inst_results}
\end{table}

\subsection{Possible Improvements for Vision Encoding}
Due to time and resource constraints, we were only able to equip the Jasper model with a basic image encoding capability. 
Initially, stage 4 was envisioned as a fundamental visual-language alignment training phase, with a potential stage 5 involving contrastive learning utilizing a Visual Question Answering (VQA) dataset. 
Additionally, we observed oscillatory behavior in our loss function during stage 4.
Overall, there is considerable room for enhancement in the multimodal training.

\section{Conclusion}

In this paper, we present the distillation-based training procedure for the Jasper model. 
We have designed three loss functions to distill multiple large teacher embedding models into a student embedding model from diverse perspectives. 
Subsequently, we utilized a MRL-based training method to reduce the vector dimensionality of the student model. Experimental results on the MTEB demonstrate that our Jasper model achieves state-of-the-art performance at the 2B parameter scale and exhibits comparable results to other top-ranked embedding models with 7B parameters.
Future work will further explore the alignment between multiple modalities.

\newpage
\newpage
\bibliography{custom}

\begin{thebibliography}{19}
\expandafter\ifx\csname natexlab\endcsname\relax\def\natexlab#1{#1}\fi

\bibitem[{Agarwal et~al.(2024)Agarwal, SK, Pancha, Hazra, Xu, and Rosenberg}]{search_www}
Prabhat Agarwal, Minhazul~Islam SK, Nikil Pancha, Kurchi~Subhra Hazra, Jiajing Xu, and Chuck Rosenberg. 2024.
\newblock \href {https://doi.org/10.1145/3589335.3648309} {Omnisearchsage: Multi-task multi-entity embeddings for pinterest search}.
\newblock In \emph{Companion Proceedings of the {ACM} on Web Conference 2024, {WWW} 2024, Singapore, Singapore, May 13-17, 2024}, pages 121--130. {ACM}.

\bibitem[{Alabdulmohsin et~al.(2024)Alabdulmohsin, Zhai, Kolesnikov, and Beyer}]{alabdulmohsin2024gettingvitshapescaling}
Ibrahim Alabdulmohsin, Xiaohua Zhai, Alexander Kolesnikov, and Lucas Beyer. 2024.
\newblock \href {http://arxiv.org/abs/2305.13035} {Getting vit in shape: Scaling laws for compute-optimal model design}.

\bibitem[{Gao et~al.(2023)Gao, Xiong, Gao, Jia, Pan, Bi, Dai, Sun, Guo, Wang, and Wang}]{rag}
Yunfan Gao, Yun Xiong, Xinyu Gao, Kangxiang Jia, Jinliu Pan, Yuxi Bi, Yi~Dai, Jiawei Sun, Qianyu Guo, Meng Wang, and Haofen Wang. 2023.
\newblock \href {https://doi.org/10.48550/ARXIV.2312.10997} {Retrieval-augmented generation for large language models: {A} survey}.
\newblock \emph{CoRR}, abs/2312.10997.

\bibitem[{Gu et~al.(2024)Gu, Zhang, Zhou, Yu, Xing, Wang, Cao, Jia, Zhang, Wang, Hu, Zhang, Li, Liang, Zhao, Ao, Liu, Feng, and Liu}]{gu2024infinitymmscalingmultimodalperformance}
Shuhao Gu, Jialing Zhang, Siyuan Zhou, Kevin Yu, Zhaohu Xing, Liangdong Wang, Zhou Cao, Jintao Jia, Zhuoyi Zhang, Yixuan Wang, Zhenchong Hu, Bo-Wen Zhang, Jijie Li, Dong Liang, Yingli Zhao, Yulong Ao, Yaoqi Liu, Fangxiang Feng, and Guang Liu. 2024.
\newblock \href {http://arxiv.org/abs/2410.18558} {Infinity-mm: Scaling multimodal performance with large-scale and high-quality instruction data}.

\bibitem[{Hofst{\"{a}}tter et~al.(2021)Hofst{\"{a}}tter, Lin, Yang, Lin, and Hanbury}]{distill_colbert}
Sebastian Hofst{\"{a}}tter, Sheng{-}Chieh Lin, Jheng{-}Hong Yang, Jimmy Lin, and Allan Hanbury. 2021.
\newblock \href {https://doi.org/10.1145/3404835.3462891} {Efficiently teaching an effective dense retriever with balanced topic aware sampling}.
\newblock In \emph{{SIGIR} '21: The 44th International {ACM} {SIGIR} Conference on Research and Development in Information Retrieval, Virtual Event, Canada, July 11-15, 2021}, pages 113--122. {ACM}.

\bibitem[{Kashyap et~al.(2024)Kashyap, Nguyen, Schlegel, Winkler, Ng, and Poria}]{embedding_representations}
Abhinav~Ramesh Kashyap, Thanh{-}Tung Nguyen, Viktor Schlegel, Stefan Winkler, See{-}Kiong Ng, and Soujanya Poria. 2024.
\newblock \href {https://aclanthology.org/2024.eacl-long.104} {A comprehensive survey of sentence representations: From the {BERT} epoch to the {CHATGPT} era and beyond}.
\newblock In \emph{Proceedings of the 18th Conference of the European Chapter of the Association for Computational Linguistics, {EACL} 2024 - Volume 1: Long Papers, St. Julian's, Malta, March 17-22, 2024}, pages 1738--1751. Association for Computational Linguistics.

\bibitem[{Kusupati et~al.(2024)Kusupati, Bhatt, Rege, Wallingford, Sinha, Ramanujan, Howard-Snyder, Chen, Kakade, Jain, and Farhadi}]{kusupati2024matryoshkarepresentationlearning}
Aditya Kusupati, Gantavya Bhatt, Aniket Rege, Matthew Wallingford, Aditya Sinha, Vivek Ramanujan, William Howard-Snyder, Kaifeng Chen, Sham Kakade, Prateek Jain, and Ali Farhadi. 2024.
\newblock \href {http://arxiv.org/abs/2205.13147} {Matryoshka representation learning}.

\bibitem[{Lee et~al.(2024)Lee, Roy, Xu, Raiman, Shoeybi, Catanzaro, and Ping}]{lee2024nv}
Chankyu Lee, Rajarshi Roy, Mengyao Xu, Jonathan Raiman, Mohammad Shoeybi, Bryan Catanzaro, and Wei Ping. 2024.
\newblock Nv-embed: Improved techniques for training llms as generalist embedding models.
\newblock \emph{arXiv preprint arXiv:2405.17428}.

\bibitem[{Li et~al.(2024)Li, Qin, Xiao, Chen, Luo, Shao, Lian, and Liu}]{li2024makingtextembeddersfewshot}
Chaofan Li, MingHao Qin, Shitao Xiao, Jianlyu Chen, Kun Luo, Yingxia Shao, Defu Lian, and Zheng Liu. 2024.
\newblock \href {http://arxiv.org/abs/2409.15700} {Making text embedders few-shot learners}.

\bibitem[{Lin et~al.(2021)Lin, Yang, and Lin}]{distll2}
Sheng{-}Chieh Lin, Jheng{-}Hong Yang, and Jimmy Lin. 2021.
\newblock \href {https://doi.org/10.18653/V1/2021.REPL4NLP-1.17} {In-batch negatives for knowledge distillation with tightly-coupled teachers for dense retrieval}.
\newblock In \emph{Proceedings of the 6th Workshop on Representation Learning for NLP, RepL4NLP@ACL-IJCNLP 2021, Online, August 6, 2021}, pages 163--173. Association for Computational Linguistics.

\bibitem[{Lozhkov et~al.(2024)Lozhkov, Ben~Allal, von Werra, and Wolf}]{lozhkov2024fineweb-edu}
Anton Lozhkov, Loubna Ben~Allal, Leandro von Werra, and Thomas Wolf. 2024.
\newblock \href {https://doi.org/10.57967/hf/2497} {Fineweb-edu: the finest collection of educational content}.

\bibitem[{Moreira et~al.(2024)Moreira, Osmulski, Xu, Ak, Schifferer, and Oldridge}]{moreira2024nv}
Gabriel de Souza~P Moreira, Radek Osmulski, Mengyao Xu, Ronay Ak, Benedikt Schifferer, and Even Oldridge. 2024.
\newblock Nv-retriever: Improving text embedding models with effective hard-negative mining.
\newblock \emph{arXiv preprint arXiv:2407.15831}.

\bibitem[{Muennighoff et~al.(2023)Muennighoff, Tazi, Magne, and Reimers}]{MTEB}
Niklas Muennighoff, Nouamane Tazi, Lo{\"{\i}}c Magne, and Nils Reimers. 2023.
\newblock \href {https://doi.org/10.18653/V1/2023.EACL-MAIN.148} {{MTEB:} massive text embedding benchmark}.
\newblock In \emph{Proceedings of the 17th Conference of the European Chapter of the Association for Computational Linguistics, {EACL} 2023, Dubrovnik, Croatia, May 2-6, 2023}, pages 2006--2029. Association for Computational Linguistics.

\bibitem[{Wang et~al.(2024)Wang, Wang, Gao, Zhang, Wu, Xu, Shi, Wang, Li, Qian, Yin, Lv, Zheng, and Huang}]{search_rag}
Xiaohua Wang, Zhenghua Wang, Xuan Gao, Feiran Zhang, Yixin Wu, Zhibo Xu, Tianyuan Shi, Zhengyuan Wang, Shizheng Li, Qi~Qian, Ruicheng Yin, Changze Lv, Xiaoqing Zheng, and Xuanjing Huang. 2024.
\newblock \href {https://aclanthology.org/2024.emnlp-main.981} {Searching for best practices in retrieval-augmented generation}.
\newblock In \emph{Proceedings of the 2024 Conference on Empirical Methods in Natural Language Processing, {EMNLP} 2024, Miami, FL, USA, November 12-16, 2024}, pages 17716--17736. Association for Computational Linguistics.

\bibitem[{Xiao et~al.(2023)Xiao, Liu, Zhang, and Muennighoff}]{bge_embedding}
Shitao Xiao, Zheng Liu, Peitian Zhang, and Niklas Muennighoff. 2023.
\newblock \href {http://arxiv.org/abs/2309.07597} {C-pack: Packaged resources to advance general chinese embedding}.

\bibitem[{Zhai et~al.(2023)Zhai, Mustafa, Kolesnikov, and Beyer}]{zhai2023sigmoidlosslanguageimage}
Xiaohua Zhai, Basil Mustafa, Alexander Kolesnikov, and Lucas Beyer. 2023.
\newblock \href {http://arxiv.org/abs/2303.15343} {Sigmoid loss for language image pre-training}.

\bibitem[{Zhao et~al.(2024{\natexlab{a}})Zhao, Liu, Ren, and Wen}]{DenseTextRetrieval}
Wayne~Xin Zhao, Jing Liu, Ruiyang Ren, and Ji{-}Rong Wen. 2024{\natexlab{a}}.
\newblock \href {https://doi.org/10.1145/3637870} {Dense text retrieval based on pretrained language models: {A} survey}.
\newblock \emph{{ACM} Trans. Inf. Syst.}, 42(4):89:1--89:60.

\bibitem[{Zhao et~al.(2024{\natexlab{b}})Zhao, Zhou, Li, Tang, Wang, Hou, Min, Zhang, Zhang, Dong, Du, Yang, Chen, Chen, Jiang, Ren, Li, Tang, Liu, Liu, Nie, and Wen}]{llm}
Wayne~Xin Zhao, Kun Zhou, Junyi Li, Tianyi Tang, Xiaolei Wang, Yupeng Hou, Yingqian Min, Beichen Zhang, Junjie Zhang, Zican Dong, Yifan Du, Chen Yang, Yushuo Chen, Zhipeng Chen, Jinhao Jiang, Ruiyang Ren, Yifan Li, Xinyu Tang, Zikang Liu, Peiyu Liu, Jian-Yun Nie, and Ji-Rong Wen. 2024{\natexlab{b}}.
\newblock \href {http://arxiv.org/abs/2303.18223} {A survey of large language models}.

\bibitem[{Zhou et~al.(2024)Zhou, Liu, Xiao, Zhao, and Xiong}]{search_multimodal}
Junjie Zhou, Zheng Liu, Shitao Xiao, Bo~Zhao, and Yongping Xiong. 2024.
\newblock \href {https://doi.org/10.18653/V1/2024.ACL-LONG.175} {{VISTA:} visualized text embedding for universal multi-modal retrieval}.
\newblock In \emph{Proceedings of the 62nd Annual Meeting of the Association for Computational Linguistics (Volume 1: Long Papers), {ACL} 2024, Bangkok, Thailand, August 11-16, 2024}, pages 3185--3200. Association for Computational Linguistics.

\end{thebibliography}


\begin{thebibliography}{80}
\expandafter\ifx\csname natexlab\endcsname\relax\def\natexlab#1{#1}\fi

\bibitem[{Aggarwal and Zhai(2012)}]{aggarwal2012survey}
Charu~C Aggarwal and ChengXiang Zhai. 2012.
\newblock A survey of text clustering algorithms.
\newblock In \emph{Mining text data}, pages 77--128. Springer.

\bibitem[{Agirre et~al.(2015)Agirre, Banea, Cardie, Cer, Diab, Gonzalez-Agirre,
  Guo, Lopez-Gazpio, Maritxalar, Mihalcea et~al.}]{agirre2015semeval}
Eneko Agirre, Carmen Banea, Claire Cardie, Daniel Cer, Mona Diab, Aitor
  Gonzalez-Agirre, Weiwei Guo, Inigo Lopez-Gazpio, Montse Maritxalar, Rada
  Mihalcea, et~al. 2015.
\newblock Semeval-2015 task 2: Semantic textual similarity, english, spanish
  and pilot on interpretability.
\newblock In \emph{Proceedings of the 9th international workshop on semantic
  evaluation (SemEval 2015)}, pages 252--263.

\bibitem[{Agirre et~al.(2014)Agirre, Banea, Cardie, Cer, Diab, Gonzalez-Agirre,
  Guo, Mihalcea, Rigau, and Wiebe}]{agirre2014semeval}
Eneko Agirre, Carmen Banea, Claire Cardie, Daniel~M Cer, Mona~T Diab, Aitor
  Gonzalez-Agirre, Weiwei Guo, Rada Mihalcea, German Rigau, and Janyce Wiebe.
  2014.
\newblock Semeval-2014 task 10: Multilingual semantic textual similarity.
\newblock In \emph{SemEval@ COLING}, pages 81--91.

\bibitem[{Agirre et~al.(2016)Agirre, Banea, Cer, Diab, Gonzalez~Agirre,
  Mihalcea, Rigau~Claramunt, and Wiebe}]{agirre2016semeval}
Eneko Agirre, Carmen Banea, Daniel Cer, Mona Diab, Aitor Gonzalez~Agirre, Rada
  Mihalcea, German Rigau~Claramunt, and Janyce Wiebe. 2016.
\newblock Semeval-2016 task 1: Semantic textual similarity, monolingual and
  cross-lingual evaluation.
\newblock In \emph{SemEval-2016. 10th International Workshop on Semantic
  Evaluation; 2016 Jun 16-17; San Diego, CA. Stroudsburg (PA): ACL; 2016. p.
  497-511.} ACL (Association for Computational Linguistics).

\bibitem[{Agirre et~al.(2012)Agirre, Cer, Diab, and
  Gonzalez-Agirre}]{agirre2012semeval}
Eneko Agirre, Daniel Cer, Mona Diab, and Aitor Gonzalez-Agirre. 2012.
\newblock Semeval-2012 task 6: A pilot on semantic textual similarity.
\newblock In \emph{* SEM 2012: The First Joint Conference on Lexical and
  Computational Semantics--Volume 1: Proceedings of the main conference and the
  shared task, and Volume 2: Proceedings of the Sixth International Workshop on
  Semantic Evaluation (SemEval 2012)}, pages 385--393.

\bibitem[{Agirre et~al.(2013)Agirre, Cer, Diab, Gonzalez-Agirre, and
  Guo}]{agirre2013sem}
Eneko Agirre, Daniel Cer, Mona Diab, Aitor Gonzalez-Agirre, and Weiwei Guo.
  2013.
\newblock * sem 2013 shared task: Semantic textual similarity.
\newblock In \emph{Second joint conference on lexical and computational
  semantics (* SEM), volume 1: proceedings of the Main conference and the
  shared task: semantic textual similarity}, pages 32--43.

\bibitem[{Allal et~al.(2023)Allal, Li, Kocetkov, Mou, Akiki, Ferrandis,
  Muennighoff, Mishra, Gu, Dey et~al.}]{allal2023santacoder}
Loubna~Ben Allal, Raymond Li, Denis Kocetkov, Chenghao Mou, Christopher Akiki,
  Carlos~Munoz Ferrandis, Niklas Muennighoff, Mayank Mishra, Alex Gu, Manan
  Dey, et~al. 2023.
\newblock Santacoder: don't reach for the stars!
\newblock \emph{arXiv preprint arXiv:2301.03988}.

\bibitem[{Andonian et~al.(2021)Andonian, Anthony, Biderman, Black, Gali, Gao,
  Hallahan, Levy-Kramer, Leahy, Nestler, Parker, Pieler, Purohit, Songz, Wang,
  and Weinbach}]{gpt-neox}
Alex Andonian, Quentin Anthony, Stella Biderman, Sid Black, Preetham Gali, Leo
  Gao, Eric Hallahan, Josh Levy-Kramer, Connor Leahy, Lucas Nestler, Kip
  Parker, Michael Pieler, Shivanshu Purohit, Tri Songz, Phil Wang, and Samuel
  Weinbach. 2021.
\newblock \href {http://github.com/eleutherai/gpt-neox} {{GPT-NeoX}: Large
  scale autoregressive language modeling in pytorch}.

\bibitem[{Angelov(2020)}]{angelov2020top2vec}
Dimo Angelov. 2020.
\newblock Top2vec: Distributed representations of topics.
\newblock \emph{arXiv preprint arXiv:2008.09470}.

\bibitem[{Asai et~al.(2020)Asai, Kasai, Clark, Lee, Choi, and
  Hajishirzi}]{asai2020xor}
Akari Asai, Jungo Kasai, Jonathan~H Clark, Kenton Lee, Eunsol Choi, and
  Hannaneh Hajishirzi. 2020.
\newblock Xor qa: Cross-lingual open-retrieval question answering.
\newblock \emph{arXiv preprint arXiv:2010.11856}.

\bibitem[{Beltagy et~al.(2019)Beltagy, Lo, and Cohan}]{beltagy2019scibert}
Iz~Beltagy, Kyle Lo, and Arman Cohan. 2019.
\newblock Scibert: A pretrained language model for scientific text.
\newblock \emph{arXiv preprint arXiv:1903.10676}.

\bibitem[{Borgeaud et~al.(2022)Borgeaud, Mensch, Hoffmann, Cai, Rutherford,
  Millican, Van Den~Driessche, Lespiau, Damoc, Clark
  et~al.}]{borgeaud2022improving}
Sebastian Borgeaud, Arthur Mensch, Jordan Hoffmann, Trevor Cai, Eliza
  Rutherford, Katie Millican, George~Bm Van Den~Driessche, Jean-Baptiste
  Lespiau, Bogdan Damoc, Aidan Clark, et~al. 2022.
\newblock Improving language models by retrieving from trillions of tokens.
\newblock In \emph{International Conference on Machine Learning}, pages
  2206--2240. PMLR.

\bibitem[{Carvalho et~al.(2018)Carvalho, Cad{\`e}ne, Picard, Soulier, Thome,
  and Cord}]{carvalho2018cross}
Micael Carvalho, R{\'e}mi Cad{\`e}ne, David Picard, Laure Soulier, Nicolas
  Thome, and Matthieu Cord. 2018.
\newblock Cross-modal retrieval in the cooking context: Learning semantic
  text-image embeddings.
\newblock In \emph{The 41st International ACM SIGIR Conference on Research \&
  Development in Information Retrieval}, pages 35--44.

\bibitem[{Casanueva et~al.(2020)Casanueva, Temčinas, Gerz, Henderson, and
  Vulić}]{casanueva2020banking77}
Iñigo Casanueva, Tadas Temčinas, Daniela Gerz, Matthew Henderson, and Ivan
  Vulić. 2020.
\newblock \href {https://doi.org/10.48550/ARXIV.2003.04807} {Efficient intent
  detection with dual sentence encoders}.

\bibitem[{Chowdhery et~al.(2022)Chowdhery, Narang, Devlin, Bosma, Mishra,
  Roberts, Barham, Chung, Sutton, Gehrmann et~al.}]{chowdhery2022palm}
Aakanksha Chowdhery, Sharan Narang, Jacob Devlin, Maarten Bosma, Gaurav Mishra,
  Adam Roberts, Paul Barham, Hyung~Won Chung, Charles Sutton, Sebastian
  Gehrmann, et~al. 2022.
\newblock Palm: Scaling language modeling with pathways.
\newblock \emph{arXiv preprint arXiv:2204.02311}.

\bibitem[{Clark et~al.(2020)Clark, Choi, Collins, Garrette, Kwiatkowski,
  Nikolaev, and Palomaki}]{clark2020tydi}
Jonathan~H Clark, Eunsol Choi, Michael Collins, Dan Garrette, Tom Kwiatkowski,
  Vitaly Nikolaev, and Jennimaria Palomaki. 2020.
\newblock Tydi qa: A benchmark for information-seeking question answering in
  typologically diverse languages.
\newblock \emph{Transactions of the Association for Computational Linguistics},
  8:454--470.

\bibitem[{Cohan et~al.(2020{\natexlab{a}})Cohan, Feldman, Beltagy, Downey, and
  Weld}]{cohan2020specter}
Arman Cohan, Sergey Feldman, Iz~Beltagy, Doug Downey, and Daniel~S Weld.
  2020{\natexlab{a}}.
\newblock Specter: Document-level representation learning using
  citation-informed transformers.
\newblock \emph{arXiv preprint arXiv:2004.07180}.

\bibitem[{Cohan et~al.(2020{\natexlab{b}})Cohan, Feldman, Beltagy, Downey, and
  Weld}]{cohan2020scidocs}
Arman Cohan, Sergey Feldman, Iz~Beltagy, Doug Downey, and Daniel~S. Weld.
  2020{\natexlab{b}}.
\newblock \href {https://doi.org/10.48550/ARXIV.2004.07180} {Specter:
  Document-level representation learning using citation-informed transformers}.

\bibitem[{Conneau and Kiela(2018)}]{conneau2018senteval}
Alexis Conneau and Douwe Kiela. 2018.
\newblock Senteval: An evaluation toolkit for universal sentence
  representations.
\newblock \emph{arXiv preprint arXiv:1803.05449}.

\bibitem[{Conneau et~al.(2017)Conneau, Kiela, Schwenk, Barrault, and
  Bordes}]{conneau2017supervised}
Alexis Conneau, Douwe Kiela, Holger Schwenk, Loic Barrault, and Antoine Bordes.
  2017.
\newblock Supervised learning of universal sentence representations from
  natural language inference data.
\newblock \emph{arXiv preprint arXiv:1705.02364}.

\bibitem[{Devlin et~al.(2018)Devlin, Chang, Lee, and
  Toutanova}]{devlin2018bert}
Jacob Devlin, Ming-Wei Chang, Kenton Lee, and Kristina Toutanova. 2018.
\newblock Bert: Pre-training of deep bidirectional transformers for language
  understanding.
\newblock \emph{arXiv preprint arXiv:1810.04805}.

\bibitem[{Fabbri et~al.(2020)Fabbri, Kryściński, McCann, Xiong, Socher, and
  Radev}]{fabbri2020summeval}
Alexander~R. Fabbri, Wojciech Kryściński, Bryan McCann, Caiming Xiong,
  Richard Socher, and Dragomir Radev. 2020.
\newblock \href {https://doi.org/10.48550/ARXIV.2007.12626} {Summeval:
  Re-evaluating summarization evaluation}.

\bibitem[{Feng et~al.(2020)Feng, Yang, Cer, Arivazhagan, and
  Wang}]{feng2020language}
Fangxiaoyu Feng, Yinfei Yang, Daniel Cer, Naveen Arivazhagan, and Wei Wang.
  2020.
\newblock Language-agnostic bert sentence embedding.
\newblock \emph{arXiv preprint arXiv:2007.01852}.

\bibitem[{FitzGerald et~al.(2022)FitzGerald, Hench, Peris, Mackie, Rottmann,
  Sanchez, Nash, Urbach, Kakarala, Singh, Ranganath, Crist, Britan, Leeuwis,
  Tur, and Natarajan}]{fitzgerald2022massive}
Jack FitzGerald, Christopher Hench, Charith Peris, Scott Mackie, Kay Rottmann,
  Ana Sanchez, Aaron Nash, Liam Urbach, Vishesh Kakarala, Richa Singh, Swetha
  Ranganath, Laurie Crist, Misha Britan, Wouter Leeuwis, Gokhan Tur, and Prem
  Natarajan. 2022.
\newblock \href {https://doi.org/10.48550/ARXIV.2204.08582} {Massive: A
  1m-example multilingual natural language understanding dataset with 51
  typologically-diverse languages}.

\bibitem[{Gao et~al.(2021{\natexlab{a}})Gao, Tow, Biderman, Black, DiPofi,
  Foster, Golding, Hsu, McDonell, Muennighoff et~al.}]{gao2021framework}
Leo Gao, Jonathan Tow, Stella Biderman, Sid Black, Anthony DiPofi, Charles
  Foster, Laurence Golding, Jeffrey Hsu, Kyle McDonell, Niklas Muennighoff,
  et~al. 2021{\natexlab{a}}.
\newblock A framework for few-shot language model evaluation.
\newblock \emph{Version v0. 0.1. Sept}.

\bibitem[{Gao and Callan(2021)}]{gao2021unsupervised}
Luyu Gao and Jamie Callan. 2021.
\newblock Unsupervised corpus aware language model pre-training for dense
  passage retrieval.
\newblock \emph{arXiv preprint arXiv:2108.05540}.

\bibitem[{Gao et~al.(2021{\natexlab{b}})Gao, Yao, and Chen}]{gao2021simcse}
Tianyu Gao, Xingcheng Yao, and Danqi Chen. 2021{\natexlab{b}}.
\newblock Simcse: Simple contrastive learning of sentence embeddings.
\newblock \emph{arXiv preprint arXiv:2104.08821}.

\bibitem[{Geigle et~al.(2021)Geigle, Reimers, Rücklé, and
  Gurevych}]{geigle2021clustering}
Gregor Geigle, Nils Reimers, Andreas Rücklé, and Iryna Gurevych. 2021.
\newblock \href {https://doi.org/10.48550/ARXIV.2104.07081} {Tweac: Transformer
  with extendable qa agent classifiers}.

\bibitem[{Heffernan et~al.(2022)Heffernan, {\c{C}}elebi, and
  Schwenk}]{heffernan2022bitext}
Kevin Heffernan, Onur {\c{C}}elebi, and Holger Schwenk. 2022.
\newblock Bitext mining using distilled sentence representations for
  low-resource languages.
\newblock \emph{arXiv preprint arXiv:2205.12654}.

\bibitem[{Hochreiter and Schmidhuber(1997)}]{hochreiter1997long}
Sepp Hochreiter and J{\"u}rgen Schmidhuber. 1997.
\newblock Long short-term memory.
\newblock \emph{Neural computation}, 9(8):1735--1780.

\bibitem[{Huang et~al.(2020)Huang, Sharma, Sun, Xia, Zhang, Pronin,
  Padmanabhan, Ottaviano, and Yang}]{huang2020embedding}
Jui-Ting Huang, Ashish Sharma, Shuying Sun, Li~Xia, David Zhang, Philip Pronin,
  Janani Padmanabhan, Giuseppe Ottaviano, and Linjun Yang. 2020.
\newblock Embedding-based retrieval in facebook search.
\newblock In \emph{Proceedings of the 26th ACM SIGKDD International Conference
  on Knowledge Discovery \& Data Mining}, pages 2553--2561.

\bibitem[{Husain et~al.(2019)Husain, Wu, Gazit, Allamanis, and
  Brockschmidt}]{husain2019codesearchnet}
Hamel Husain, Ho-Hsiang Wu, Tiferet Gazit, Miltiadis Allamanis, and Marc
  Brockschmidt. 2019.
\newblock Codesearchnet challenge: Evaluating the state of semantic code
  search.
\newblock \emph{arXiv preprint arXiv:1909.09436}.

\bibitem[{Izacard et~al.(2021)Izacard, Caron, Hosseini, Riedel, Bojanowski,
  Joulin, and Grave}]{izacard2021towards}
Gautier Izacard, Mathilde Caron, Lucas Hosseini, Sebastian Riedel, Piotr
  Bojanowski, Armand Joulin, and Edouard Grave. 2021.
\newblock Towards unsupervised dense information retrieval with contrastive
  learning.
\newblock \emph{arXiv preprint arXiv:2112.09118}.

\bibitem[{Komninos and Manandhar(2016)}]{komninos2016dependency}
Alexandros Komninos and Suresh Manandhar. 2016.
\newblock Dependency based embeddings for sentence classification tasks.
\newblock In \emph{Proceedings of the 2016 conference of the North American
  chapter of the association for computational linguistics: human language
  technologies}, pages 1490--1500.

\bibitem[{Lan et~al.(2017)Lan, Qiu, He, and Xu}]{lan2017sentential}
Wuwei Lan, Siyu Qiu, Hua He, and Wei Xu. 2017.
\newblock \href {http://aclweb.org/anthology/D17-1127} {A continuously growing
  dataset of sentential paraphrases}.
\newblock In \emph{Proceedings of The 2017 Conference on Empirical Methods on
  Natural Language Processing (EMNLP)}, pages 1235--1245. Association for
  Computational Linguistics.

\bibitem[{Lhoest et~al.(2021)Lhoest, del Moral, Jernite, Thakur, von Platen,
  Patil, Chaumond, Drame, Plu, Tunstall et~al.}]{lhoest2021datasets}
Quentin Lhoest, Albert~Villanova del Moral, Yacine Jernite, Abhishek Thakur,
  Patrick von Platen, Suraj Patil, Julien Chaumond, Mariama Drame, Julien Plu,
  Lewis Tunstall, et~al. 2021.
\newblock Datasets: A community library for natural language processing.
\newblock \emph{arXiv preprint arXiv:2109.02846}.

\bibitem[{Li et~al.(2020)Li, Arora, Chen, Gupta, Gupta, and
  Mehdad}]{li2020mtop}
Haoran Li, Abhinav Arora, Shuohui Chen, Anchit Gupta, Sonal Gupta, and Yashar
  Mehdad. 2020.
\newblock \href {https://doi.org/10.48550/ARXIV.2008.09335} {Mtop: A
  comprehensive multilingual task-oriented semantic parsing benchmark}.

\bibitem[{Liu et~al.(2018)Liu, Wang, Leng, and Zhai}]{liu2018linkso}
Xueqing Liu, Chi Wang, Yue Leng, and ChengXiang Zhai. 2018.
\newblock Linkso: a dataset for learning to retrieve similar question answer
  pairs on software development forums.
\newblock In \emph{Proceedings of the 4th ACM SIGSOFT International Workshop on
  NLP for Software Engineering}, pages 2--5.

\bibitem[{Maas et~al.(2011)Maas, Daly, Pham, Huang, Ng, and
  Potts}]{maas2011imdb}
Andrew~L. Maas, Raymond~E. Daly, Peter~T. Pham, Dan Huang, Andrew~Y. Ng, and
  Christopher Potts. 2011.
\newblock \href {https://aclanthology.org/P11-1015} {Learning word vectors for
  sentiment analysis}.
\newblock In \emph{Proceedings of the 49th Annual Meeting of the Association
  for Computational Linguistics: Human Language Technologies}, pages 142--150,
  Portland, Oregon, USA. Association for Computational Linguistics.

\bibitem[{McAuley and Leskovec(2013)}]{mcauley2013amazon}
Julian McAuley and Jure Leskovec. 2013.
\newblock \href {https://doi.org/10.1145/2507157.2507163} {Hidden factors and
  hidden topics: Understanding rating dimensions with review text}.
\newblock RecSys '13, New York, NY, USA. Association for Computing Machinery.

\bibitem[{Muennighoff(2020)}]{muennighoff2020vilio}
Niklas Muennighoff. 2020.
\newblock Vilio: State-of-the-art visio-linguistic models applied to hateful
  memes.
\newblock \emph{arXiv preprint arXiv:2012.07788}.

\bibitem[{Muennighoff(2022)}]{muennighoff2022sgpt}
Niklas Muennighoff. 2022.
\newblock Sgpt: Gpt sentence embeddings for semantic search.
\newblock \emph{arXiv preprint arXiv:2202.08904}.

\bibitem[{Muennighoff et~al.(2022)Muennighoff, Wang, Sutawika, Roberts,
  Biderman, Scao, Bari, Shen, Yong, Schoelkopf
  et~al.}]{muennighoff2022crosslingual}
Niklas Muennighoff, Thomas Wang, Lintang Sutawika, Adam Roberts, Stella
  Biderman, Teven~Le Scao, M~Saiful Bari, Sheng Shen, Zheng-Xin Yong, Hailey
  Schoelkopf, et~al. 2022.
\newblock Crosslingual generalization through multitask finetuning.
\newblock \emph{arXiv preprint arXiv:2211.01786}.

\bibitem[{Nayak(2019)}]{nayak2021google}
Pandu Nayak. 2019.
\newblock \href
  {https://blog.google/products/search/search-language-understanding-bert/}
  {Understanding searches better than ever before}.

\bibitem[{Neelakantan et~al.(2022)Neelakantan, Xu, Puri, Radford, Han, Tworek,
  Yuan, Tezak, Kim, Hallacy et~al.}]{neelakantan2022text}
Arvind Neelakantan, Tao Xu, Raul Puri, Alec Radford, Jesse~Michael Han, Jerry
  Tworek, Qiming Yuan, Nikolas Tezak, Jong~Wook Kim, Chris Hallacy, et~al.
  2022.
\newblock Text and code embeddings by contrastive pre-training.
\newblock \emph{arXiv preprint arXiv:2201.10005}.

\bibitem[{Ni et~al.(2021{\natexlab{a}})Ni, {\'A}brego, Constant, Ma, Hall, Cer,
  and Yang}]{ni2021sentence}
Jianmo Ni, Gustavo~Hern{\'a}ndez {\'A}brego, Noah Constant, Ji~Ma, Keith~B
  Hall, Daniel Cer, and Yinfei Yang. 2021{\natexlab{a}}.
\newblock Sentence-t5: Scalable sentence encoders from pre-trained text-to-text
  models.
\newblock \emph{arXiv preprint arXiv:2108.08877}.

\bibitem[{Ni et~al.(2021{\natexlab{b}})Ni, Qu, Lu, Dai, {\'A}brego, Ma, Zhao,
  Luan, Hall, Chang et~al.}]{ni2021large}
Jianmo Ni, Chen Qu, Jing Lu, Zhuyun Dai, Gustavo~Hern{\'a}ndez {\'A}brego,
  Ji~Ma, Vincent~Y Zhao, Yi~Luan, Keith~B Hall, Ming-Wei Chang, et~al.
  2021{\natexlab{b}}.
\newblock Large dual encoders are generalizable retrievers.
\newblock \emph{arXiv preprint arXiv:2112.07899}.

\bibitem[{Nichol et~al.(2021)Nichol, Dhariwal, Ramesh, Shyam, Mishkin, McGrew,
  Sutskever, and Chen}]{nichol2021glide}
Alex Nichol, Prafulla Dhariwal, Aditya Ramesh, Pranav Shyam, Pamela Mishkin,
  Bob McGrew, Ilya Sutskever, and Mark Chen. 2021.
\newblock Glide: Towards photorealistic image generation and editing with
  text-guided diffusion models.
\newblock \emph{arXiv preprint arXiv:2112.10741}.

\bibitem[{O'Neill et~al.(2021)O'Neill, Rozenshtein, Kiryo, Kubota, and
  Bollegala}]{oneill2021amazoncounterfactual}
James O'Neill, Polina Rozenshtein, Ryuichi Kiryo, Motoko Kubota, and Danushka
  Bollegala. 2021.
\newblock \href {https://doi.org/10.48550/ARXIV.2104.06893} {I wish i would
  have loved this one, but i didn't -- a multilingual dataset for
  counterfactual detection in product reviews}.

\bibitem[{Pedregosa et~al.(2011)Pedregosa, Varoquaux, Gramfort, Michel,
  Thirion, Grisel, Blondel, Prettenhofer, Weiss, Dubourg, Vanderplas, Passos,
  Cournapeau, Brucher, Perrot, and Duchesnay}]{scikit-learn}
F.~Pedregosa, G.~Varoquaux, A.~Gramfort, V.~Michel, B.~Thirion, O.~Grisel,
  M.~Blondel, P.~Prettenhofer, R.~Weiss, V.~Dubourg, J.~Vanderplas, A.~Passos,
  D.~Cournapeau, M.~Brucher, M.~Perrot, and E.~Duchesnay. 2011.
\newblock Scikit-learn: Machine learning in {P}ython.
\newblock \emph{Journal of Machine Learning Research}, 12:2825--2830.

\bibitem[{Pennington et~al.(2014)Pennington, Socher, and
  Manning}]{pennington2014glove}
Jeffrey Pennington, Richard Socher, and Christopher~D Manning. 2014.
\newblock Glove: Global vectors for word representation.
\newblock In \emph{Proceedings of the 2014 conference on empirical methods in
  natural language processing (EMNLP)}, pages 1532--1543.

\bibitem[{Radford et~al.(2019)Radford, Wu, Child, Luan, Amodei, Sutskever
  et~al.}]{radford2019language}
Alec Radford, Jeffrey Wu, Rewon Child, David Luan, Dario Amodei, Ilya
  Sutskever, et~al. 2019.
\newblock Language models are unsupervised multitask learners.
\newblock \emph{OpenAI blog}, 1(8):9.

\bibitem[{Raffel et~al.(2020)Raffel, Shazeer, Roberts, Lee, Narang, Matena,
  Zhou, Li, Liu et~al.}]{raffel2020exploring}
Colin Raffel, Noam Shazeer, Adam Roberts, Katherine Lee, Sharan Narang, Michael
  Matena, Yanqi Zhou, Wei Li, Peter~J Liu, et~al. 2020.
\newblock Exploring the limits of transfer learning with a unified text-to-text
  transformer.
\newblock \emph{J. Mach. Learn. Res.}, 21(140):1--67.

\bibitem[{Reimers et~al.(2016)Reimers, Beyer, and Gurevych}]{reimers2016task}
Nils Reimers, Philip Beyer, and Iryna Gurevych. 2016.
\newblock Task-oriented intrinsic evaluation of semantic textual similarity.
\newblock In \emph{Proceedings of COLING 2016, the 26th International
  Conference on Computational Linguistics: Technical Papers}, pages 87--96.

\bibitem[{Reimers and Gurevych(2019)}]{reimers2019sentence}
Nils Reimers and Iryna Gurevych. 2019.
\newblock Sentence-bert: Sentence embeddings using siamese bert-networks.
\newblock \emph{arXiv preprint arXiv:1908.10084}.

\bibitem[{Research()}]{tatoeba}
Facebook Research.
\newblock \href
  {https://github.com/facebookresearch/LASER/tree/main/data/tatoeba/v1}
  {Tatoeba multilingual test set}.

\bibitem[{Rosenberg and Hirschberg(2007)}]{vmeasure}
Andrew Rosenberg and Julia Hirschberg. 2007.
\newblock V-measure: A conditional entropy-based external cluster evaluation
  measure.
\newblock pages 410--420.

\bibitem[{Saharia et~al.(2022)Saharia, Chan, Saxena, Li, Whang, Denton,
  Ghasemipour, Ayan, Mahdavi, Lopes et~al.}]{saharia2022photorealistic}
Chitwan Saharia, William Chan, Saurabh Saxena, Lala Li, Jay Whang, Emily
  Denton, Seyed Kamyar~Seyed Ghasemipour, Burcu~Karagol Ayan, S~Sara Mahdavi,
  Rapha~Gontijo Lopes, et~al. 2022.
\newblock Photorealistic text-to-image diffusion models with deep language
  understanding.
\newblock \emph{arXiv preprint arXiv:2205.11487}.

\bibitem[{Saravia et~al.(2018)Saravia, Liu, Huang, Wu, and
  Chen}]{saravia2018emotion}
Elvis Saravia, Hsien-Chi~Toby Liu, Yen-Hao Huang, Junlin Wu, and Yi-Shin Chen.
  2018.
\newblock \href {https://doi.org/10.18653/v1/D18-1404} {{CARER}: Contextualized
  affect representations for emotion recognition}.
\newblock In \emph{Proceedings of the 2018 Conference on Empirical Methods in
  Natural Language Processing}, pages 3687--3697, Brussels, Belgium.
  Association for Computational Linguistics.

\bibitem[{Scao et~al.(2022)Scao, Fan, Akiki, Pavlick, Ili{\'c}, Hesslow,
  Castagn{\'e}, Luccioni, Yvon, Gall{\'e} et~al.}]{scao2022bloom}
Teven~Le Scao, Angela Fan, Christopher Akiki, Ellie Pavlick, Suzana Ili{\'c},
  Daniel Hesslow, Roman Castagn{\'e}, Alexandra~Sasha Luccioni, Fran{\c{c}}ois
  Yvon, Matthias Gall{\'e}, et~al. 2022.
\newblock Bloom: A 176b-parameter open-access multilingual language model.
\newblock \emph{arXiv preprint arXiv:2211.05100}.

\bibitem[{Shah et~al.(2018)Shah, Lei, Moschitti, Romeo, and
  Nakov}]{shah2018adversarial}
Darsh Shah, Tao Lei, Alessandro Moschitti, Salvatore Romeo, and Preslav Nakov.
  2018.
\newblock \href {https://doi.org/10.18653/v1/D18-1131} {Adversarial domain
  adaptation for duplicate question detection}.
\newblock In \emph{Proceedings of the 2018 Conference on Empirical Methods in
  Natural Language Processing}, pages 1056--1063, Brussels, Belgium.
  Association for Computational Linguistics.

\bibitem[{Song et~al.(2020)Song, Tan, Qin, Lu, and Liu}]{song2020mpnet}
Kaitao Song, Xu~Tan, Tao Qin, Jianfeng Lu, and Tie-Yan Liu. 2020.
\newblock Mpnet: Masked and permuted pre-training for language understanding.
\newblock \emph{Advances in Neural Information Processing Systems},
  33:16857--16867.

\bibitem[{Srivastava et~al.(2022)Srivastava, Rastogi, Rao, Shoeb, Abid, Fisch,
  Brown, Santoro, Gupta, Garriga-Alonso et~al.}]{srivastava2022beyond}
Aarohi Srivastava, Abhinav Rastogi, Abhishek Rao, Abu Awal~Md Shoeb, Abubakar
  Abid, Adam Fisch, Adam~R Brown, Adam Santoro, Aditya Gupta, Adri{\`a}
  Garriga-Alonso, et~al. 2022.
\newblock Beyond the imitation game: Quantifying and extrapolating the
  capabilities of language models.
\newblock \emph{arXiv preprint arXiv:2206.04615}.

\bibitem[{Tan and Bansal(2019)}]{tan2019lxmert}
Hao Tan and Mohit Bansal. 2019.
\newblock Lxmert: Learning cross-modality encoder representations from
  transformers.
\newblock \emph{arXiv preprint arXiv:1908.07490}.

\bibitem[{Thakur et~al.(2021)Thakur, Reimers, Rücklé, Srivastava, and
  Gurevych}]{beir}
Nandan Thakur, Nils Reimers, Andreas Rücklé, Abhishek Srivastava, and Iryna
  Gurevych. 2021.
\newblock \href {https://doi.org/10.48550/ARXIV.2104.08663} {Beir: A
  heterogenous benchmark for zero-shot evaluation of information retrieval
  models}.

\bibitem[{Vaswani et~al.(2017)Vaswani, Shazeer, Parmar, Uszkoreit, Jones,
  Gomez, Kaiser, and Polosukhin}]{vaswani2017attention}
Ashish Vaswani, Noam Shazeer, Niki Parmar, Jakob Uszkoreit, Llion Jones,
  Aidan~N Gomez, {\L}ukasz Kaiser, and Illia Polosukhin. 2017.
\newblock Attention is all you need.
\newblock \emph{Advances in neural information processing systems}, 30.

\bibitem[{Wang et~al.(2019)Wang, Pruksachatkun, Nangia, Singh, Michael, Hill,
  Levy, and Bowman}]{wang2019superglue}
Alex Wang, Yada Pruksachatkun, Nikita Nangia, Amanpreet Singh, Julian Michael,
  Felix Hill, Omer Levy, and Samuel Bowman. 2019.
\newblock Superglue: A stickier benchmark for general-purpose language
  understanding systems.
\newblock \emph{Advances in neural information processing systems}, 32.

\bibitem[{Wang et~al.(2018)Wang, Singh, Michael, Hill, Levy, and
  Bowman}]{wang2018glue}
Alex Wang, Amanpreet Singh, Julian Michael, Felix Hill, Omer Levy, and Samuel~R
  Bowman. 2018.
\newblock Glue: A multi-task benchmark and analysis platform for natural
  language understanding.
\newblock \emph{arXiv preprint arXiv:1804.07461}.

\bibitem[{Wang and Komatsuzaki(2021)}]{gpt-j}
Ben Wang and Aran Komatsuzaki. 2021.
\newblock {GPT-J-6B: A 6 Billion Parameter Autoregressive Language Model}.
\newblock \url{https://github.com/kingoflolz/mesh-transformer-jax}.

\bibitem[{Wang et~al.(2021)Wang, Reimers, and Gurevych}]{wang2021tsdae}
Kexin Wang, Nils Reimers, and Iryna Gurevych. 2021.
\newblock Tsdae: Using transformer-based sequential denoising auto-encoder for
  unsupervised sentence embedding learning.
\newblock \emph{arXiv preprint arXiv:2104.06979}.

\bibitem[{Wang et~al.(2020)Wang, Wei, Dong, Bao, Yang, and
  Zhou}]{wang2020minilm}
Wenhui Wang, Furu Wei, Li~Dong, Hangbo Bao, Nan Yang, and Ming Zhou. 2020.
\newblock Minilm: Deep self-attention distillation for task-agnostic
  compression of pre-trained transformers.
\newblock \emph{Advances in Neural Information Processing Systems},
  33:5776--5788.

\bibitem[{Weinbach et~al.(2022)Weinbach, Bellagente, Eichenberg, Dai, Baldock,
  Nanda, Deiseroth, Oostermeijer, Teufel, and Cruz-Salinas}]{weinbach2022m}
Samuel Weinbach, Marco Bellagente, Constantin Eichenberg, Andrew Dai, Robert
  Baldock, Souradeep Nanda, Bj{\"o}rn Deiseroth, Koen Oostermeijer, Hannah
  Teufel, and Andres~Felipe Cruz-Salinas. 2022.
\newblock M-vader: A model for diffusion with multimodal context.
\newblock \emph{arXiv preprint arXiv:2212.02936}.

\bibitem[{Wolf et~al.(2020)Wolf, Debut, Sanh, Chaumond, Delangue, Moi, Cistac,
  Rault, Louf, Funtowicz et~al.}]{wolf2020transformers}
Thomas Wolf, Lysandre Debut, Victor Sanh, Julien Chaumond, Clement Delangue,
  Anthony Moi, Pierric Cistac, Tim Rault, R{\'e}mi Louf, Morgan Funtowicz,
  et~al. 2020.
\newblock Transformers: State-of-the-art natural language processing.
\newblock In \emph{Proceedings of the 2020 conference on empirical methods in
  natural language processing: system demonstrations}, pages 38--45.

\bibitem[{Wu et~al.(2020)Wu, Qiao, Chen, Wu, Qi, Lian, Liu, Xie, Gao, Wu
  et~al.}]{wu2020mind}
Fangzhao Wu, Ying Qiao, Jiun-Hung Chen, Chuhan Wu, Tao Qi, Jianxun Lian,
  Danyang Liu, Xing Xie, Jianfeng Gao, Winnie Wu, et~al. 2020.
\newblock Mind: A large-scale dataset for news recommendation.
\newblock In \emph{Proceedings of the 58th Annual Meeting of the Association
  for Computational Linguistics}, pages 3597--3606.

\bibitem[{Xu et~al.(2015)Xu, Callison-Burch, and Dolan}]{xu2015semeval}
Wei Xu, Chris Callison-Burch, and William~B Dolan. 2015.
\newblock Semeval-2015 task 1: Paraphrase and semantic similarity in twitter
  (pit).
\newblock In \emph{Proceedings of the 9th international workshop on semantic
  evaluation (SemEval 2015)}, pages 1--11.

\bibitem[{Zhang et~al.(2022)Zhang, Thakur, Ogundepo, Kamalloo, Alfonso-Hermelo,
  Li, Liu, Rezagholizadeh, and Lin}]{zhang2022making}
Xinyu Zhang, Nandan Thakur, Odunayo Ogundepo, Ehsan Kamalloo, David
  Alfonso-Hermelo, Xiaoguang Li, Qun Liu, Mehdi Rezagholizadeh, and Jimmy Lin.
  2022.
\newblock Making a miracl: Multilingual information retrieval across a
  continuum of languages.
\newblock \emph{arXiv preprint arXiv:2210.09984}.

\bibitem[{Zhu et~al.(2021)Zhu, Li, Li, and Oduola}]{zhu2021bing}
Jeffrey Zhu, Mingqin Li, Jason Li, and Cassandra Oduola. 2021.
\newblock \href
  {https://blogs.bing.com/Engineering-Blog/october-2021/Bing-delivers-more-contextualized-search-using-quantized-transformer-inference-on-NVIDIA-GPUs-in-Azu}
  {Bing delivers more contextualized search using quantized transformer
  inference on nvidia gpus in azure}.

\bibitem[{Zweigenbaum et~al.(2016)Zweigenbaum, Sharoff, and
  Rapp}]{zweigenbaum2016bucc1}
Pierre Zweigenbaum, Serge Sharoff, and Reinhard Rapp. 2016.
\newblock Towards preparation of the second bucc shared task: Detecting
  parallel sentences in comparable corpora.
\newblock In \emph{Proceedings of the Ninth Workshop on Building and Using
  Comparable Corpora. European Language Resources Association (ELRA), Portoroz,
  Slovenia}, pages 38--43.

\bibitem[{Zweigenbaum et~al.(2017)Zweigenbaum, Sharoff, and
  Rapp}]{zweigenbaum2017bucc2}
Pierre Zweigenbaum, Serge Sharoff, and Reinhard Rapp. 2017.
\newblock Overview of the second bucc shared task: Spotting parallel sentences
  in comparable corpora.
\newblock In \emph{Proceedings of the 10th Workshop on Building and Using
  Comparable Corpora}, pages 60--67.

\bibitem[{Zweigenbaum et~al.(2018)Zweigenbaum, Sharoff, and
  Rapp}]{zweigenbaum2018bucc3}
Pierre Zweigenbaum, Serge Sharoff, and Reinhard Rapp. 2018.
\newblock Overview of the third bucc shared task: Spotting parallel sentences
  in comparable corpora.
\newblock In \emph{Proceedings of 11th workshop on building and using
  comparable corpora}, pages 39--42.

\end{thebibliography}
\bibliographystyle{acl_natbib}

\end{document}